\documentstyle[aps,pra,prabib,psfig,epsfig]{revtex}

\renewcommand{\widetext}{}

\renewcommand{\narrowtext}{}

\begin{document}

\newlength{\insideFig}

\title{Contribution of the screened self-energy to the Lamb shift of
  quasidegenerate states}

\author{\'Eric-Olivier Le~Bigot\cite{bylineeol}
\and Paul Indelicato\cite{bylinepi}
\and Vladimir M. Shabaev\cite{bylinevms}
}
\address{
Laboratoire Kastler-Brossel, Case 74,
\'Ecole Normale Sup\'erieure et Universit\'e P. et M. Curie\\
Unit\'e Mixte de Recherche du CNRS n$^\circ$ C8552\\
4, pl.~Jussieu, 75252 Paris CEDEX 05, France}

\date{\today}

\maketitle

\begin{abstract}
  Expressions for the effective Quantum Electrodynamics (QED)
  Hamiltonian due to self-energy screening (self-energy correction to
  the electron-electron interaction) are presented.  We use the method
  of the two-time Green's function, which handles quasidegenerate
  atomic states.  From these expression one can evaluate energy
  corrections to, e.g., $1s2p\,^3P_1$ and $1s2p\,^1P_1$ in helium and
  two-electron ions, to all orders in $Z\alpha$.
\end{abstract}

\pacs{31.30.Jv, 31.15.-p, 31.30.-i, 31.15.Ar}
\draft

\narrowtext

In the last ten years, experiments in the spectroscopy of
helium~\cite{sansonetti90,lichten91,marin94,dorrer97,eikema97} have
become two orders of magnitude more precise than the best theoretical
energy level calculations available (see, e.g., Refs.\ 
\cite{drake98,drake2000} and references therein). Several experiments
are now focusing on Helium and heliumlike ions
$1s2p\,^3P_J$ fine structure
\cite{minardi99,storry2000,castillega2000,myers99,myers2000}, with the
aim of providing a new determination of the fine structure constant
and of checking higher-order effects in the calculations.  In this
case the theory is again a limiting factor.  In this context a direct
determination of all $\alpha^2$ contributions to all order in
$Z\alpha$ is necessary to improve reliability and accuracy of
theoretical calculations ($\alpha$ being the fine structure constant,
and $Z$ the charge of the nucleus).

A difficulty in the study of the $(1s2p_{1/2})_1$ and $(1s2p_{3/2})_1$
levels is that they are quasidegenerate for low and middle $Z$ ions
\cite{artemyev2000}; this precludes the use of the Gell-Man--Low and
Sucher method \cite{gellmann51,sucher57} to evaluate QED energy shifts
of atomic levels. In fact, this method has two important drawbacks: it
does not handle quasidegenerate energy levels, and it leads to a
difficult renormalization procedure when applied to degenerate states.
(The latter problem has only been tackled up to second-order in
$\alpha$\ \cite{braun80,braun84}.)

We use the method of the two-time Green's function
\cite{shabaev90,shabaev94,shabaev94b}, rigorously derived from QED
(for the most detailed description of this method, see
\cite{shabaev2000}).  To the best of our knowledge, only the method
recently proposed by Lindgren \cite{lindgren2000}, closely modeled to
multireference-state Many-body perturbation techniques, is designed to
work for quasidegenerate states.

We evaluate the contribution of the screened self-energy diagrams
\begin{equation}\label{eq:selfEnergyDiag}
\parbox{36mm}{ %

\psfig{figure=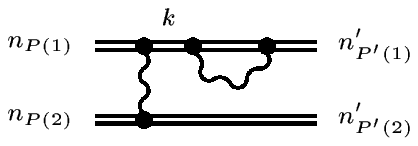}

}
\quad
\parbox{30mm}{ %

\psfig{figure=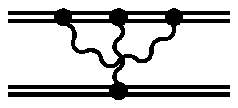}

}
\end{equation}
to quasidegenerate energy levels in heliumlike ions. Our results
can be easily extended to ions with more than two electrons along
lines similar to those found in \cite{shabaev93}.

First approximate evaluations of the contribution of these
  diagrams for isolated states in two- and three-electron ions were
  performed in Refs.\ 
  \cite{indelicato91,blundell93a,lindgren93,chen95}.  Accurate
  calculations from the first principles of QED were accomplished in
  Refs.\  \cite{yerokhin97b,persson96a,sunnergren98} for the
  \emph{ground state} of \emph{heliumlike} ions and in
  Refs.\  \cite{yerokhin98,yerokhin99} for the $2s$ and $2p_{1/2}$
  states of \emph{lithiumlike} ions.  The other two $\alpha^2$
  corrections to the electron-electron interaction have also been
  calculated for \emph{isolated} states in two- and
  three-electron ions: the \emph{vacuum-polarization screening}
  \cite{artemyev2000,persson96a,sunnergren98,artemyev97,artemyev99},
  and the \emph{two-photon exchange} diagrams
  \cite{blundell93,lindgren95,mohr2000,yerokhin2000b}.  In
  \cite{artemyev2000}, the vacuum polarization screening for
  quasidegenerate states of heliumlike ions was evaluated as well.
  Some results for the direct contribution of the self-energy
  correction to the Coulomb interaction are also
  available~\cite{indelicato91,indelicato2000}.

As depicted in diagrams (\ref{eq:selfEnergyDiag}), the interaction
between the two electrons through photons is treated perturbatively.
On the contrary, the binding to the nucleus is included
\emph{non-perturbatively} in the method we use, since the
corresponding coupling constant is $Z\alpha$. Such a
treatment is obviously
mandatory for highly-charged ions. Furthermore, it allows one to compare
non-perturbative (in $Z\alpha$) results to (semi-)analytic expansions
in $Z\alpha$ (see \cite{pachucki98b} for a review).

We derive the effective (finite-sized) matrix
hamiltonian $H$, whose eigenvalues give the contribution of QED
to a group of energy levels~\cite{shabaev93}.  The diagonal entries of
the hamiltonian that we evaluate correctly reproduce previous
expressions of the screened self-energy,
while the new, non-diagonal entries that we derive allow one to obtain
a second-order QED correction to \emph{quasidegenerate} or
\emph{degenerate} energy
levels.

Relativistic units $\hbar=c=1$ are used throughout this paper.

\label{sec:hamiltonian}

If we have $s$ quasidegenerate energy levels
$E_{1\ldots s}^{(0)}$, the effective hamiltonian
$H$ is an $s\times s$ matrix restricted to these
levels~\cite{shabaev93}. Let us introduce some notations in order to
express this hamiltonian. The second-order contribution $H^{(2)}$ to this hamiltonian $H =
H^{(0)} + H^{(1)} +
H^{(2)} + \ldots$ is constructed from a projection matrix
${P}$ and an energy matrix
${K}$~\cite{shabaev93}:
\begin{eqnarray}
\nonumber
H^{(2)} &=&
{K}^{(2)}
-{\frac{1}{2}}\{ {P}^{(1)},{K}^{(1)}\}
-{\frac{1}{2}}\{ {P}^{(2)},{K}^{(0)}\}
\\
\label{eq:effHamOrig}
&&
+\frac{3}{8}\{ [{P}^{(1)}]{}^2,{K}^{(0)}\}
+\frac{1}{4}{P}^{(1)} {K}^{(0)}
{P}^{(1)},
\end{eqnarray}
where the notation $\{,\}$ represents the usual anticommutator, and
where the superscripts indicate the number of photons of the diagrams
that contribute to each term of the perturbative expansion
${P} = {P}^{(0)} + {P}^{(1)}
+\ldots$ and ${K} = {K}^{(0)} +
{K} ^{(1)} + \ldots$; the $s\times s$ matrices
${P}$ and ${K}$, which are defined
as~\cite{shabaev94b}:
\begin{mathletters}
\label{eq:defProjAndEnergyMatrices}
\begin{eqnarray}
{P} &\equiv& \frac{1}{2\pi i}
 {\displaystyle \oint_{\Gamma}\!}
  d{E}\,{g(E)} \\
{K} &\equiv& \frac{1}{2\pi i}
 {\displaystyle \oint_{\Gamma}\!}
  d{E}\,{E\, g(E)},
\end{eqnarray}
\end{mathletters}%
where $g(E)$ is the $s\times s$ matrix restriction of the
Green's function to the $s$ unperturbed atomic levels under
consideration, and where $\Gamma$ is a contour that encloses each of
the Dirac atomic energy levels with a positive orientation~\cite{shabaev93}.

We directly evaluate the hamiltonian matrix elements of
Eq.\ (\ref{eq:effHamOrig}) between states of \emph{different} energies
$E_{n}^{(0)}$ and $E_{n'}^{(0)}$, and put them in a
form that readily displays the limiting case of \emph{identical}
energies; we checked by a direct calculation of the diagonal matrix
elements that they can be obtained from non-diagonal elements
$H^{(2)}_{nn'}$ by taking the formal limit
$E_{n}^{(0)}\rightarrow E_{n'}^{(0)}$. \emph{All}
the subsequent derivations of $H^{(2)}_{nn'}$ will thus be
done with $E_{n}^{(0)} \ne E_{n'}^{(0)}$.

The first diagram of (\ref{eq:selfEnergyDiag}) appears only in
the second-order matrices ${K}^{(2)}$ and
${P}^{(2)}$ in Eq.\ (\ref{eq:effHamOrig}). As usual, we must
calculate a \emph{reducible} and an \emph{irreducible}
contribution; as can be seen in subsequent calculations, it turns out
that the correct extension of these notions to quasidegenerate states
is the following: in the first diagram of Eq.\ (\ref{eq:selfEnergyDiag}), the
contribution of intermediate electrons with a Dirac energy
$\varepsilon_k$ such that $\varepsilon_k+\varepsilon_{n'_{P'(2)}}$
coincides with one of the $s$ energy levels under consideration and
must be \emph{separated out} from the contribution of the other
intermediate electron states; the first contribution (called
\emph{reducible}) requires a different mathematical treatment
from that of the second contribution (called \emph{irreducible}).

\label{sec:intermElectrons}

Thus, the \emph{irreducible} contribution is obtained by summing
over almost all electron states $k$ in the first diagram of
Eq.\ (\ref{eq:selfEnergyDiag}); we first show that it is sufficient to remove
only \emph{one} state $k$ from the sum over states in the first
diagram of Eq.\ (\ref{eq:selfEnergyDiag}). We see that an intermediate energy
$\varepsilon_k+\varepsilon_{n'_{P'(2)}}$ can coincide with an
unperturbed atomic levels $E_{1\ldots s}^{(0)}$ only if the
electron $k$ has the \emph{same principal quantum number} as the
electron $n'_{P'(1)}$ on the other side of the self-energy, because
otherwise the total energy $\varepsilon_k+\varepsilon_{n'_{P'(2)}}$
would lie largely out of the range spanned by the unperturbed
quasidegenerate energy levels located around $E_{n'}^{(0)} =
\varepsilon_{n'_{P'(1)}} +\varepsilon_{n'_{P'(2)}}$.

There is an additional selection on the electrons $k$ to be removed:
since the total angular momentum, its projection, and parity are
conserved by the self-energy operator $\Sigma$ [Eq.\ (\ref{eq:defSigma})
below], as can be seen by integrating over angles using standard
techniques~\cite{grant70}, the contribution of electrons $k$ that do
not share the same quantum numbers $(\kappa,m)$ as the electron
$n'_{P'(1)}$ in the first diagram of Eq.\ (\ref{eq:selfEnergyDiag}) is exactly
\emph{zero}.

We denote the individual electrons of a state $n$ by $n_1$ and $n_2$,
in an order which is arbitrary but that must remain fixed. With these
notations, our evaluation of the \emph{irreducible} part of the
first diagram of (\ref{eq:selfEnergyDiag}) to the effective
hamiltonian (\ref{eq:effHamOrig}) takes a simple form and reads (Dirac
energies are still denoted by $\varepsilon_k$): \widetext
\begin{eqnarray}
\nonumber
H^{\mbox{\scriptsize scr. SE, irr.}}_{nn'}
&
=
&
\sum_{P,P'} {(-1)^{PP'}}\Bigg(
\sum_{k \ne n_{P(1)}}
\mbox{$\langle n_{P(1)}|$}\Sigma(\varepsilon_{n_{P(1)}})\mbox{$|k\rangle$}
\frac{1}{\varepsilon_{n_{P(1)}} - \varepsilon_{k}}
\mbox{$\langle k n_{P(2)}|$}I(\varepsilon_{n_{P(1)}}-\varepsilon_{n'_{P'(1)}})\mbox{$|m'_1
  m'_2\rangle$}
\\
\nonumber
&&
+
\sum_{k \ne n'_{P'(1)}}
\mbox{$\langle n_{P(1)} n_{P(2)}|$}I(\varepsilon_{n_{P(1)}}-\varepsilon_{n'_{P'(1)}})\mbox{$|n'_{P'(1)} n'_{P'(2)}\rangle$}
\frac{1}{ \varepsilon_{n'_{P'(1)}} - \varepsilon_{k}}
\mbox{$\langle k|$}\Sigma(\varepsilon_{n'_{P'(1)}})\mbox{$|n'_{P'(1)}\rangle$}
\Bigg) 
\\
\label{eq:effHamSSEIrr}
&&
+
{\cal {O}}[\alpha^2 (E_{n'}^{(0)}-E_{n}^{(0)})],
\end{eqnarray}
\narrowtext
\noindent%
where ${(-1)^{PP'}}$ is the signature of the permutation
$P\circ P'$ ($P$ and $P'$ are permutations of $\{1,2\}$.), where the
sum over $k$ is over (almost) all possible intermediate Dirac
states, and where the photon exchange and the self-energy of diagrams
(\ref{eq:selfEnergyDiag}) are represented by the following usual
operators~\cite{yerokhin99}:
\begin{eqnarray}
\nonumber
\mbox{$\langle ab|$}I(\omega)\mbox{$|cd\rangle$}
 &\equiv&
e^2
\int \! d^{3}\bbox{x}_1\,\int \! d^{3}\bbox{x}_2\,
[\psi_a^{\dagger}(\bbox{x}_1)\alpha^{\mu}\psi_c(\bbox{x}_1)]
\\
\label{eq:defI}
&&\times
[\psi_b^{\dagger}(\bbox{x}_2)\alpha^{\nu}\psi_d(\bbox{x}_2)]
D_{\mu\nu}(\omega; \bbox{x}_1-\bbox{x}_2)
\\
\label{eq:defSigma}
\mbox{$\langle a|$}\Sigma(p)\mbox{$|b\rangle$}
&\equiv&
\frac{1}{2\pi i} \int \! d\omega \,
\sum_{k}\frac{\mbox{$\langle ak|$}I(\omega)\mbox{$|kb\rangle$}}
{\varepsilon_k(1-i0) - (p-\omega)},
\end{eqnarray}
in which $e$ is the charge of the electron, $\alpha^{\mu}\equiv
\left(1,\bbox{\alpha}\right)$ are the Dirac matrices, and where $\psi$
denotes a Dirac spinor; the photon propagator $D$ is given in the
Feynman gauge by
\begin{equation}
D_{\nu\nu'}(\omega; \bbox{r})
 \equiv 
g_{\nu\nu'} \frac{\exp\left(i|\bbox{r}|\sqrt{\omega^2-\mu^2+i
      0}
  \right)}
{4\pi|\bbox{r}|},
\end{equation}
where $\mu$ is a small photon mass that eventually tends to zero, and
where the square root branch is chosen such as to yield a decreasing
exponential for large real-valued energies $\omega$.  

The last term in Eq.\ (\ref{eq:effHamSSEIrr}) represents a contribution of
  order $\alpha^2$ which is multiplied by a factor that tends to zero
  as $E_{n'}^{(0)}-E_{n}^{(0)} \rightarrow 0$. It can be shown
    (see Ref.\  \cite{shabaev2000}) that such a term does not
    contribute to order $\alpha^2$ and that it can therefore be
    omitted.

We note that result (\ref{eq:effHamSSEIrr}) readily yields diagonal
elements by taking the (formal) limit $E_{n}^{(0)} -
E_{n'}^{(0)} \rightarrow 0$.

The hamiltonian (\ref{eq:effHamOrig}) contains the contribution of
many \emph{first}-order diagrams through the operators
${P}^{(1)}$ and ${K}^{(1)}$. We must
consider here the contribution of the photon exchange and of the
self-energy
\begin{equation}
\label{eq:photExchAndSE}
\parbox{17mm}{ %

\psfig{figure=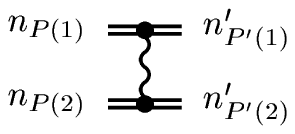}

}
\qquad\qquad
\parbox{27mm}{ %

\psfig{figure=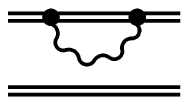}

};
\end{equation}
their contribution to Eq.\ (\ref{eq:effHamOrig}) \emph{cancels} a part of
the \emph{reducible} screened self-energy.  We thus evaluate in the
following the contribution of both diagrams of Eq.\ (\ref{eq:photExchAndSE}) to
the terms $
-{\frac{1}{2}}\{ {P}^{(1)},{K}^{(1)}\}
+\frac{3}{8}\{ [{P}^{(1)}]{}^2,{K}^{(0)}\}
+\frac{1}{4}{P}^{(1)} {K}^{(0)}
{P}^{(1)} $ of the effective hamiltonian.

The energy and projection matrices ${K}$ and
${P}$ of Eq.\ (\ref{eq:defProjAndEnergyMatrices}) have been calculated
for the photon-exchange diagram in~\cite{artemyev2000}; this allows
one to evaluate any integral due to the \emph{photon exchange}
that appears in the effective hamiltonian (\ref{eq:effHamOrig}).

  In order to derive the contribution of the \emph{one-electron}
  self-energy, let us show that the evaluation of the self-energy
  contributions to the hamiltonian (\ref{eq:effHamOrig}) boils down to
  the calculation of contour integrals of the form
\begin{equation}\label{eq:SENeeded}
\frac{1}{2\pi i}
 {\displaystyle \oint_{\Gamma_n}\!}
  d{E}\,{g^{{\rm {SE}}}_{nn}(E)}
\quad{\rm {and}}\quad
\frac{1}{2\pi i}
 {\displaystyle \oint_{\Gamma_n}\!}
  d{E}\,{E\, g^{{\rm {SE}}}_{nn}(E)}
\end{equation}
where $g^{{\rm {SE}}}_{nn}(E)$ are diagonal elements of the self-energy
Green's function; in other words, the contour $\Gamma$ that
surrounds \emph{all} the levels in Eq.\ (\ref{eq:defProjAndEnergyMatrices})
can be replaced by the contour that surrounds $E_{n}^{(0)}$
\emph{only}, and \emph{non-diagonal} elements of the
self-energy Green's function are not relevant. The contour integrals
of Eq.\ (\ref{eq:SENeeded}) have both been evaluated in~\cite{yerokhin99}, so
that no further quantity is required in order to obtain the
self-energy contribution to the hamiltonian~(\ref{eq:effHamOrig}).

Let us prove the above statements. As mentioned before, angular
momentum conservations constrain the self-energy operator
$\Sigma$ to be zero between states with different angular quantum numbers
$(\kappa,m)$; and since the atomic levels we consider have the same
principal quantum number (they are quasidegenerate), the self-energy
Green's matrix is diagonal:
\begin{equation}
g^{{\rm {SE}}}_{nn'}(E)  = 0 \quad {\rm {if\ }}n\ne n',
\end{equation}
where $n$ and $n'$ are the \emph{sets} of quantum numbers of two
of the $s$ levels under consideration.

Furthermore, the Green's function $g^{{\rm {SE}}}_{nn}(E)$ has only
\emph{one} pole inside the integration contour $\Gamma$, namely at
$E=E_{n}^{(0)}$. Therefore, integrating over the full contour
$\Gamma$ in the hamiltonian (\ref{eq:effHamOrig}) amounts to
integrate over the contour $\Gamma_n$ that surrounds only
$E_{n}^{(0)}$, since the Green's function is analytic inside
the contours that encircle the other energies.

We thus see that the contribution of the self-energy to
Eq.\ (\ref{eq:effHamOrig}) depends only on contour integrals of the form
(\ref{eq:SENeeded}), which are known analytically~\cite{yerokhin99}.

With the help of some published analytical formulas, we obtain the
following contribution of the photon exchange (see Eqs.\ (27) and (28)
in \cite{artemyev2000}) and of the self-energy (see Eqs.\ (36) and (37)
in \cite{yerokhin99}) to  the effective
hamiltonian~(\ref{eq:effHamOrig}):
\widetext
\begin{eqnarray}
\label{eq:photExchAndSEContrib}
\nonumber
\lefteqn{
-\sum_{P,P'} {(-1)^{PP'}}
\Bigg\{
\frac{1}{4}
\Bigg[
\left(
\mbox{$\langle n_{P(1)}|$}\Sigma'(\varepsilon_{n_{P(1)}})\mbox{$|n_{P(1)}\rangle$}
+
\mbox{$\langle n'_{P(1)}|$}\Sigma'(\varepsilon_{n'_{P(1)}})\mbox{$|n'_{P(1)}\rangle$}
\right)
}
&&
\\
\nonumber
&&
\qquad
\times
\left(
\mbox{$\langle n_{P(1)} n_{P(2)}|$}I(\Delta_1)\mbox{$|n'_{P'(1)} n'_{P'(2)}\rangle$}
+
\mbox{$\langle n_{P(1)} n_{P(2)}|$}I(\Delta_2)\mbox{$|n'_{P'(1)} n'_{P'(2)}\rangle$}
\right)
\Bigg]
\\
\nonumber
&&
+\frac{1}{2}
\Bigg[
\left(
\mbox{$\langle n_{P(1)}|$}\Sigma(\varepsilon_{n_{P(1)}})\mbox{$|n_{P(1)}\rangle$}
+
\mbox{$\langle n'_{P(1)}|$}\Sigma(\varepsilon_{n'_{P(1)}})\mbox{$|n'_{P(1)}\rangle$}
\right)
\\
&&
\times
\frac{1}{2\pi i}
\int \! d\omega \,
\mbox{$\langle n_{P(1)} n_{P(2)}|$}I(\omega)\mbox{$|n'_{P'(1)} n'_{P'(2)}\rangle$}
\left(
\frac{1}{(\omega+\Delta_1-i0) (\omega-\Delta_2-i0) }
+
\frac{1}{(\omega+\Delta_2-i0) (\omega-\Delta_1-i0) }
\right)
\Bigg]
\Bigg\}
,
\end{eqnarray}%
\narrowtext%
\noindent%
where $\Sigma'$ represents the derivative of the self-energy
operator (\ref{eq:defSigma}) with respect to the energy that flows in
it, and where the two possible energies for the photon in the
photon-exchange diagram are $\Delta_1 \equiv
\varepsilon_{n_{P(1)}} - \varepsilon_{n'_{P'(1)}}$ and
$\Delta_2 \equiv \varepsilon_{n_{P(2)}} -
\varepsilon_{n'_{P'(2)}}$.

As seen above, the \emph{reducible} part of the \emph{first}
diagram of Eq.\ (\ref{eq:selfEnergyDiag}) represents the contribution of an
intermediate electron $k={n'_{P'(1)}}$. (For the
\emph{second} diagram, the reducible part is similarly obtained
through an intermediate electron $k= {n_{P(1)}}$.) The
evaluation of the reducible contribution follows steps similar to
those used for the irreducible part.
The contribution of diagrams (\ref{eq:photExchAndSE}) to the effective
hamiltonian $H^{(2)}$, which is given in
Eq.\ (\ref{eq:photExchAndSEContrib}), \emph{cancels} a few terms of the
contribution of the reducible diagram, as for diagonal matrix elements
\cite{yerokhin99}; the total \emph{reducible} contribution to
Eq.\ (\ref{eq:effHamOrig}) is then found to be: \widetext
\begin{eqnarray}
\nonumber
H^{\mbox{\scriptsize scr. SE, red.}}_{nn'}
&=&
\sum_{P,P'} {(-1)^{PP'}} {\frac{1}{2}}
\bigg[
\left.\partial_{p}\right|_{\varepsilon_{n_{P(1)}}}
\left(
\mbox{$\langle n_{P(1)}|$}\Sigma(p)\mbox{$|n_{P(1)}\rangle$}
\mbox{$\langle n_{P(1)} n_{P(2)}|$}I(p-\varepsilon_{n'_{P'(1)}})\mbox{$|n'_{P'(1)}
  n'_{P'(2)}\rangle$}
\right)
\\
\nonumber
&&
\quad
+
\left.\partial_{p'}\right|_{\varepsilon_{n'_{P'(1)}}}
\left(
\mbox{$\langle n_{P(1)} n_{P(2)}|$}I(\varepsilon_{n_{P(1)}} - p')\mbox{$|n'_{P'(1)}
  n'_{P'(2)}\rangle$}
\mbox{$\langle n'_{P'(1)}|$}\Sigma(p')\mbox{$|n'_{P'(1)}\rangle$}
\right)
\bigg]\\
\label{eq:redContribFinal}
&&
+ {\cal {O}}{[\alpha^2 (E_{n'}^{(0)}-E_{n}^{(0)})]}
,
\end{eqnarray}
\narrowtext
\noindent%
where $\left.\partial_x\right|_{x_0}$ represents the derivative with
respect to $x$ at the point $x_0$.

  For the vertex diagram [second diagram of (\ref{eq:selfEnergyDiag})],
  the two-time Green's function method yields the following contribution to
  (\ref{eq:effHamOrig}): \widetext
\begin{eqnarray}
\nonumber
\lefteqn{
H_{nn'}^{{\rm {vertex}}}
=
\sum_{P,P'} {(-1)^{PP'}} \sum_{i_1,i_2} \mbox{$\langle i_1 n_{P(2)}|$}I(\varepsilon_{n_{P(1)}} - \varepsilon_{n'_{P'(1)}})\mbox{$|i_2
  n'_{P'(2)}\rangle$}
}&&
\\
\label{eq:vertexContrib}
&&
\quad\times
\frac{i}{2\pi} \int \! d\omega \, \frac{\mbox{$\langle n_{P(1)}
    i_2|$}I(\omega)\mbox{$|i_1 n'_{P'(1)}\rangle$}}{
[\varepsilon_{i_1}(1-i0)-(\varepsilon_{n_{P(1)}} - \omega)]
[\varepsilon_{i_2}(1-i0)-(\varepsilon_{n'_{P'(1)}} - \omega)]
}
+{\cal {O}}[\alpha^2 (E_{n'}^{(0)}-E_{n}^{(0)})],
\end{eqnarray}
\narrowtext\noindent%
with the same notations as before; the sum is over all pairs of Dirac states.

We thus have obtained the full contribution [Eq.\ (\ref{eq:effHamSSEIrr}) +
Eq.\ (\ref{eq:redContribFinal})+Eq.\ (\ref{eq:vertexContrib})] of the screened self-energy
diagrams (\ref{eq:selfEnergyDiag}) to a finite-sized effective
hamiltonian which acts on a few atomic energy levels (in the general
case: quasidegenerate, fully degenerate or isolated); the eigenvalues
of this hamiltonian give the QED prediction for the energy levels. We
have also taken into account the contribution of the first-order
diagrams (\ref{eq:photExchAndSE}) to the second-order hamiltonian
(\ref{eq:effHamOrig}).

The results presented here extend previous derivations of the
  screened self-energy contribution to the Lamb shift, which were
restricted to the evaluation of the energy shift of an
\emph{isolated} level. The diagonal terms of the effective
hamiltonian that we have evaluated confirm previously published
results. The new, non-diagonal matrix elements of the hamiltonian that
we obtained allow one to calculate the energy shifts of
quasidegenerate levels and to extend numerical
calculations~\cite{indelicato91,yerokhin97b,persson96a,sunnergren98,yerokhin98,indelicato98}
to such levels.

 Partial support for this research has been provided by the
European Community under TMR contract number FMRX-CT97-0144
(EUROTRAPS). V.\  M.\  S.\  wishes to thank the \' Ecole Normale
Sup\' erieure for providing support during the completion of this work.

\end{document}